\documentclass[Journal]{copernicus}

\usepackage{url,lineno}
\usepackage{color}
\usepackage[T1]{fontenc}

\begin{document}

\title{Kinetic theory of information -- the dynamics of information
}

\author[1,2]{R. A. Treumann\thanks{Visiting the International Space Science Institute, Bern, Switzerland}}
\author[3]{W. Baumjohann}

\affil[1]{Department of Geophysics and Environmental Sciences, Munich University, Munich, Germany}
\affil[2]{International Space Science Institute, Bern, Switzerland}
\affil[3]{Space Research Institute, Austrian Academy of Sciences, Graz, Austria}

\runningtitle{Information Dynamics}

\runningauthor{R. A. Treumann and W. Baumjohann}

\correspondence{R. A.Treumann\\ (rudolf.treumann@geophysik.uni-muenchen.de)}

\received{ }
\revised{ }
\accepted{ }
\published{ }


\firstpage{1}

\maketitle

\paragraph*{\bf{Abstract}}
A kinetic approach to the notion of information is proposed, based on Liouville kinetic theory. The general kinetic equation for the evolution of the N-particle information $\mathcal{I}_N$ in a Hamiltonian system of large particle number $N\gg 1$ is obtained. It is shown that the $N$-particle information is strictly conserved. Defining reduced particle number information densities in phase space should be possible to obtain a kinetic equation for the ordinary one-particle information $\mathcal{I}_1\equiv \mathcal{I}$ following the Bogoliubov prescription. The kinetic equation for $\mathcal{I}$ is a kind of generalized Boltzmann equation with interaction term depending on the hierarchy of reduced informations. This term in its general form is the most general expression for the Kolmogorov entropy rate of evolution of the information.  


\section*{Introduction}

The internal information content of a physical system is its entropy $S$ which, in classical Hamiltonian systems with {one-particle (index 1) Hamiltonian $H_1(q,p)=\sum_{i} (p_i^2/2m)+\sum_{j\neq i}U(q_j)$, is a function of the particle coordinate vectors  $p,q$ in Boltzmann's 6-dimensional (3 time dependent momentum vector $p(t)$ and 3 space vector $q(t)$ coordinates, with $t$ time) one-particle $\mu$-phase space, mass $m$ and potential energy $U(q_j)$, respectively, with $i, j$ particle number.} It depends on the complete dynamics of all indistinguishable particles on the kinetic level under the action of the interparticle forces (accounted for in the potential $U$) which contribute to $H_1$ and are defined by the Hamilton equations of motion {$\dot{q}=\partial H_1/\partial p, \dot{p}=-\partial H_1/\partial q$}. This internal information is, according to Boltzmann and Shannon, given as the product of the {$\mu$-}phase space density $F_1(p,q,t)$, the {one-particle} distribution function, and its logarithm $\log F_1$, with $F_1$ satisfying the Boltzmann equation {$\partial_tF_1+[H_1,F_1]=\mathcal{C}_B$, with $[\dots,\dots]\equiv(\partial_pH_1)(\partial_qF_1)-(\partial_qH_1)(\partial_pF_1)$ the one particle Poisson bracket, and $\mathcal{C}_B$ Boltzmann's collision integral}.  

In any realistic {classical} physical system composed of $N$ subsystems (particles) of very large number $N\gg 1$, the Boltzmann equation is replaced by the Liouville equation
\begin{equation}\label{liouville}
\mathcal{L}_NF_N=0, \qquad\mathrm{with}\qquad \mathcal{L}_N=\frac{\partial}{\partial t} +[H_N, ...]
\end{equation}
{which holds in the 6$N$-dimensional Gibbs' $\Gamma$-phase space \citep[for a general reference cf., e.g.,][including a discussion of quantum systems]{balescu2000}.} The operator $\mathcal{L}_N$ is the $N$-particle Liouville operator. It contains the $N$-particle Hamiltonian $H_N(p_N,q_N)$, which in classical theory is not an operator. Here the brackets $[...,...]$ are $N$-particle Poisson's brackets. 

This Liouville equation, unlike the Boltzmann equation, is exact. It acts on the exact phase space density, which for classical point particles is defined as
\begin{equation}\label{distfunc}
F_N(q,p,t)=\prod_i^N \delta [q-q_i(t)]\delta[p-p_i(t)]
\end{equation}
All dynamics is contained in the time-dependent phase space coordinates of the $N$ particles (subsystems) $q_i(t), p_i(t)$ via Hamilton's equations. $F_N$, being a phase space (probability) density, must be normalised accordingly.  In this sense, Liouville's and Hamilton's equations are tautologies. However, the former offers the advantage of a probabilistic approach which avoids the necessity of solving for all $N$ fully dynamical Hamilton equations {$\dot{q}_N=\partial H_N/\partial p_N, \dot{p}_N=-\partial H_N/\partial q_N$}.

The  Liouville equation describes the flow of the $N$ particles through $N$-particle phase-space under the action of the $N$-particle Hamiltonian, explicitly exhibiting the conservation of particles. Its ({formally known}) solution $F_N$ though exact is rather impractical. {It requires knowledge of all $N$ exact particle orbits at all times $t$.} The Liouville equation can, however, be reduced to a one-particle Boltzmann-like equation holding in one-particle phase space (with all particles identical distinguished only by mass, charge, and energy) following a complicated procedure of reductive integration known as the BBGKY hierarchy {(after N. N. Bogoliubov, M. Born, H. S. Green, J. G. Kirkwood, and J. Yvonne) \cite[for reference cf., e.g.][]{bogoliubov1962,huang1987}. We will refer to an equivalent of this approach below in Section 3.} 

The question which interests us here concerns the nature of entropy/information on the $N$-particle level in $N$-particle {Gibbs' $\Gamma$}-phase space. Answering this question should provide an evolution theory for the information of a physical system. In other words, one may hope obtaining a Boltzmann-like {one-particle $\mu$-phase} space kinetic equation for the information.

{This question is not purely academic. Interest in the physics of information arose primarily with the advent of chaos theory in the early sixties \cite[for a collection of different approaches in different fields the reader is referred to the Santa Fe proceedings volume][]{zurek1990}. Indeed, information is one of the central quantities not only in physics but also in several other fields like communication, biology etc. all referring in one or the other sense to information theory. Information theory makes use of a physical definition of information that is exploited for instance in maximum entropy methods of data analysis as also in various applications to the determination of probability measures, chaotic behaviour, as well as quite practical problems like weather prediction in meteorology, climate research, prediction theory in the evolution of time series, in space physics in general and space weather in particular \citep[in the latter respect see, e.g., the comprehensive review in][and references therein]{balasis2013} where low-dimensional chaotic approaches have found wide application. Thermodynamics predicts that information cannot be lost. Under stationary conditions it enters thermodynamics respectively statistical mechanics being the central quantity in the first and second thermodynamic laws where it relates directly to internal energy and external work done on the system. Under non-stationary conditions its evolution is barely known while being of utmost importance. Since thermodynamics is rooted in kinetic theory it is reasonable to ask whether information cannot be subject to kinetic theory as well. } 

{So far, chaos theory of low-dimensional systems provides tools already to infer about generation of information and its effects on the system. However, most physical systems are composed of very many subsystems the interaction of which should be taken into account when considering information, inferring about its evolution, production and accumulation. Such a theory should be rooted in first physical principles, i.e. for systems consisting of many subsystems and having large numbers of degrees of freedom it should be based on Liouville theory in order to make information accessible to well known physical methods, technical treatment and providing a deeper understanding of its evolution and distribution throughout the system and sharing by the various subsystems. Since information, once generated, cannot be lost, such distribution processes are of  importance in particular in view of processes which, like information spread in societies, so far are not subject to any treatment in physics. In such systems apparently very small amounts of energy are involved causing large effects by information transfer. In the following we briefly sketch how a time-dependent \emph{physical} information theory could be developed.}  

\section*{N-particle Information}
Let us  define an equivalent $N$-particle phase-space information density $\mathcal{I}_N$ following the Boltzmann-Shannon prescription
\begin{equation}\label{inform1}
\mathcal{I}_N(F_N)= F_N \log F_N
\end{equation}
assuming that it is also normalized (for instance to Boltzmann's constant $k_B$). Via the phase space density $F_N$ it depends on the complete phase space dynamics of the  system  contained in the Hamiltonian and Hamilton's equations. The question is then, which equation does the exact $N$-particle information satisfy? 

In order to answer this question we tentatively apply the above Liouville operator $\mathcal{L}_N$ to $\mathcal{I}_N$ keeping in mind that it  applies strictly only to the $N$-particle phase space distribution function $F_N$. Before doing this we rewrite $\mathcal{I}_N$ using the definition of the exact phase space density Equation (\ref{distfunc})
\begin{equation}
\mathcal{I}_N(F_N)=\prod_i\delta(q-q_i)\delta(p-p_i)\log\prod_j\delta(q-q_j)\delta(p-p_j)
\end{equation}
the logarithm of the product becomes a sum $\sum_j\log\delta(q-q_j)\delta(p-p_j)$. Mathematically the logarithm of a distribution -- the delta function -- must be taken with care; this is a weakness in the Shannon definition on the microscopic level. Hence the expression is to be considered as a formal representation only. Formally, however, the delta functions take care for that all mixed products vanish. Hence one finds the obvious result that 
\begin{equation}\label{inform2}
\mathcal{I}_N=\sum_i^N\mathcal{I}_i \qquad \mathrm{with} \qquad \mathcal{I}_i=F_i\log F_i
\end{equation}
is the sum of all single particle informations. In other words, the $N$-particle information is additive (extensive). This is a consequence of the logarithmic dependence imposed by the Shannon prescription to which we restrict here. Other definitions as, for instance, the generalisations available in the literature \cite[for instance][]{renyi1970,wehrl1978,tsallis1988} or that given recently in \cite{treumann2014} may destroy the extensivity already on the Liouville level by adding correlations in the last expression. 

If we  apply the Liouville operator to Equations (\ref{inform1}) and (\ref{inform2}), it is easy to demonstrate for any conservative system that
\begin{equation}
\mathcal{L}_N\mathcal{I}_N=(1+\log\mathcal{I}_N)\mathcal{L}_NF_N =0
\end{equation}
which holds for any physically reasonable $\log\mathcal{I}_N\neq-1$ and because of Equation (\ref{liouville}). For this reason the exact $N$-particle information satisfies the exact $N$-particle Liouville equation
\begin{equation}\label{info-liouville}
\mathcal{L}_N\mathcal{I}_N(F_N)=0
\end{equation}
The $N$-particle information thus follows Liouville dynamics in phase space with the dynamic equations prescribed by the $N$-particle Hamiltonian function. We may note here that this is a classical and no quantum theory. Generalisation to quantum theory is by no means obvious. The nonlinear nature of the information inhibits the simple replacement of Poisson's brackets with commutators/anti-commutators.

Within Shannon theory we therefore find an exact kinetic equation for the exact classical $N$-particle Shannon information $\mathcal{I}_N$, and that this equation is the Liouville equation (\ref{info-liouville}) acting on the full $N$-particle phase space. In principle, this is an expected result.

However, like in the case of \emph{exact kinetic} particle theory, this equation is an identity and not yet an equation with that one could operate. It just says that in the exact $N$-particle phase space the $N$-particle information is conserved and behaves like a phase space information density which corresponds to an information flow through phase space. In fact this is not such a surprise. Since no averaging has been done when following the exact orbits of all particles, no overall disorder of the phase space is produced. The total information content which has been there in the total phase space volume at the beginning is still there; it is simply conserved. Information could not have gone anywhere from the total volume. 

It is again noteworthy that this conclusion is strictly valid only for the $N$-particle Shannon information. Other definitions might not reproduce Liouville's equation in its known form with vanishing right hand side. In the general case one expects that an $N$-particle diffusion term occurs on the right in this case being responsible for the dispersion of the generalised information in $N$-particle phase space and violating the extensivity of the $N$-particle information expressed in Equ. (\ref{info-liouville}). If this is the case, there must be some physical reason imposed from the outside for the definition of information at variance. Non-conservation of information then means that loss or gain of total information is attributed to this reason. One may, for instance, think of interaction with some external field which, on the global level, extracts information from the system or adds information to it. We will not consider this case here.

\section*{Hierarchy}
Physical reality as experienced in practice does, however, not take place on the microscopic level of $N$-particle phase space. The path to a practical kinetic theory of (Shannon) information is in principle prescribed by the analogy to kinetic theory. There a hierarchy of average $n=(N-j)$-particle distributions $F_n$,  with $j\in\mathsf{N}, j\leq N-1$ is prescribed as the suitably normalised  integral over the phase space coordinates of all particles $j$. This procedure, known as BBGKY hierarchy, ultimately reduces the phase space to the one-particle {$\mu$-phase space} of the now undistinguishable $N$ particles. It in this case produces the average one-particle phase space distribution function $f(q,p)\equiv F_1$ and, from Liouville's equation, the ultimate one-particle kinetic equation describing the evolution of $f(q,p) 
$ under the action of the reduced one-particle  Hamiltonian $H_1(q,p)$. This final one-particle Boltzmann-like kinetic equation contains a non-vanishing right-hand side which collects all correlations between the particles and their mutual interaction fields {\citep[cf., e.g,][for their explicit classical and quantum forms]{montgomery1964,reichl1980,balescu2000}}. In case of merely hard-core binary collisions, this term reduces further to {Boltzmann's $\mu$-space} equation which, in the total absence of any collisions and just for purely classical field interactions, becomes the zero-right hand side Vlasov equation {\citep[][Ch. 3]{klimontovich}} or, including quasilinear interactions with self-excited field fluctuations in the one-particle kinetic equation, the Fokker-Planck equation {\citep[for a rigorous and lucid derivation of the Fokker-Planck equations cf., e.g.,][]{montgomery1964}}.  

A similar procedure should go along the lines of an analogous definition of phase-space-averaged ``reduced" informations $\mathcal{I}_n$ forming a descending in $n$ chain. The philosophy behind this approach is that the Shannon information  is understood as the average of the logarithm of the distribution function itself. Thus any reduced information is given as
\begin{equation}
\mathcal{I}_n=\frac{\int\mathrm{d}^{3(n+1)}q\mathrm{d}^{3(n+1)}pF_{n+1}(q,p,t)\log F_{n+1}(q,p,t)}{\int\mathrm{d}^{3(n+1)}q\mathrm{d}^{3(n+1)}pF_{n+1}}
\end{equation}
The problem consists in finding the kinetic equation that governs the evolution of $\mathcal{I}_n$ with $n\to1$ from the exact Liouville equation for $\mathcal{I}_N$. This step is substantially more complex and less transparent than in the case of the BBGKY hierarchy of the $n$-particle distribution function which results in the Boltzmann equation. There the nonlinearity is provided by the Hamiltonian, while in the case of the information the $n$-particle information itself is intrinsically nonlinear, and care must be taken in each step when applying the reduced Hamiltonians.

Carrying through this program is a formidable task though being quite straightforward. In every step of the reduction procedure one has to take care of the reduced Liouville operator, the hierarchy equation of the former step for the reduced distribution function. Already in the BBGKY hierarchy a non-vanishing and rather complicated term on the right-hand side of the reduced kinetic equation is produced to which each reduction step adds further terms. This will also happen in the case of the reduction of the Liouville equation {(\ref{info-liouville})} for the information. In this note we refrain from performing all these steps leaving it to someone else who can explicit it. The present paper is just a perspective paper that intends to present the basic idea of constructing a viable physical theory of the dynamics of information. However, some simple arguments can be given what the reduction procedure, the information hierarchy, will lead us to.

In the case of the BBGKY hierarchy the form of the Liouville equation on the left hand side of the equation reproduces in each step with descending subsystem (particle) number $n$ from $n=N$ down to $n=1$, the wanted final form of the kinetic equation. All correlations and subsystem (particle) interactions become relegated to form an ever more complicated interaction term on the right hand side of these equations. Because the structure of the $N$-subsystem Liouville equation for the $N$-subsystem information $\mathcal{I}_N$ is identical to that of the $N$-particle Liouville equation for $F_N$, we expect that this behaviour will reproduce also in the case of the information hierarchy. We conjecture that the final one-subsystem (particle) kinetic equation for the information will be of the form
\begin{equation}
\mathcal{L}_1\mathcal{I}_1(p_1,q_1)= \mathcal{C}\{q_1,q_2 \dots q_N;p_1,p_2 \dots p_N\}
\end{equation}
where $\mathcal{L}_1\equiv\mathcal{L}=\partial_t+[H_1(p,q),\dots]$,  with $p\equiv p_1,q\equiv q_1$ {and $H_1(p,q)$ the one-particle Hamiltonian}, is the one-particle Liouville operator, and $\mathcal{C}$, as indicated by the braces $\{\dots\}$, functionally contains all coordinates and correlations. It results in the reductive procedure leading from the $N$- to the one-subsystem kinetic equation. All the dynamics that causes the evolution of the information in the interaction between the particles is contained in $\mathcal{C}$. Since the Hamiltonian also contains collisionless interactions with external and self-consistent fields, such interactions are taken care on the left-hand side of the above kinetic equation. It is, however, questionable whether neglect of the right-hand side can be justified in the case of information as is  done in Vlasov and Klimontovich \citep{klimontovich} theory. Presumably, since information (entropy) cannot be erased but can only grow, no comparably simple argument can be found for dropping $\mathcal{C}$ in the case of information unless the system is in thermodynamic equilibrium when its dynamics and evolution is obsolete. 

\section*{Discussion}
Presumably, the dynamical theory of Boltzmann-Shannon information will result in a Boltzmann-like kinetic equation for the one-subsystem information $\mathcal{I}_1$  with, however, very complicated non-vanishing correlation term on its right. It is clear that this is particular only to the Boltzmann-Shannon information. Other definitions of entropy-information found in the literature might not lead to similar kind of reductions. Having conjectured the form of the one-subsystem kinetic equation for the one-subsystem information and determined the \emph{implicit}  functional form of the correlation term on the right of the above equation, will provide a full kinetic theory of the evolution of  information in an $N$-subsystem configuration. This is most interesting for a number of obvious reasons and may apply not only to physics but also to other sciences and engineering where information generation and evolution plays an important role. Here we have restricted ourselves to physics alone.

To close this perspective article we point out that the above conjectured kinetic equation for the one-subsystem information can also be understood differently. The reduced (one-subsystem) Liouville operator on the left can be written as the total time derivative in the one-subsystem phase space. The kinetic equation then reads
\begin{equation}
\mathcal{L}_1\mathcal{I}_1(p_1,q_1)\equiv \frac{\mathrm{d}}{\mathrm{d}t}\mathcal{I}_1(t) = \mathcal{C}\{q_1(t),\dots q_N(t);p_1(t),\dots p_N(t)\}
\end{equation}
The formal solution of this version is
\begin{equation}
\mathcal{I}_1(t)=\int_{-\infty}^t \mathrm{d}t \ \mathcal{C}\{q_1(t),\dots q_N(t);p_1(t),\dots p_N(t)\}
\end{equation}
which yields the time evolution of the one-subsystem information. The difficulty here lies not only in the necessity to know the explicit functional form of $\mathcal{C}$ which follows from the hierarchy approach, it is relegated in addition to the knowledge of time dependence of the one-, two-,\ \dots subsystem trajectories in phase space, which strictly spoken implies the knowledge of the full particle dynamics in phase space which are not known a priori. Thus the above formal solution is a tautology, and one has to apply some kind of approximation like perturbation methods to treat the perturbation of an initial state. This resembles the situation encountered in kinetic theory. Nevertheless, though this is not completely satisfactory, the above form shows that the total time derivative of the one-subsystem information is determined by the functional $\mathcal{C}$. Conventionally, in the theory of chaotic interactions of a small number of particles this is taken to be the so-called Kolmogorov entropy rate $K$ {\citep{kolmogorov1958,kolmogorov1959} which originally appeared as a metric entropy rate in Kolmogorov-Arnold-Moser (KAM) theory of chaotic processes \citep[for reference cf., e.g.][]{broer2004}}. Hence, performing the derivation of the hierarchy leads in a straight way to the physical definition of the Kolmogorov $K$ entropy rate for an $N$-subsystem configuration as
\begin{equation}
K(t)\equiv \mathcal{C}\{q_1(t),\dots q_N(t);p_1(t),\dots p_N(t)\}
\end{equation}
as an implicit function of time and the number of subsystems (particles).  {For any non-stationary system $\mathcal{C}\neq 0$ for the simple reason that entropy/information can only grow.} Since $\mathcal{C}$ itself  depends implicitly on the higher order informations, the Kolmogorov entropy rate is also a functional of information. Moreover, $t=\mathcal{I}^{-1}$ is the (properly normalised) inverse functional of the information. One thus writes
\begin{equation}
{\mathrm{d}\mathcal{I}/\mathrm{d}t}=K\{\mathcal{I}\}
\end{equation}
where, for simplicity, we dropped the index 1. From here we obtain the interesting expression 
\begin{equation}
\mathrm{d}t=\mathrm{d}\,\mathcal{I}/K\{\mathcal{I}\}
\end{equation}
with its formal solution
\begin{equation}
t-t_0=\int_{\mathcal{I}_0}^\mathcal{I} \mathrm{d}\,\mathcal{I}/K\{\mathcal{I}\}
\end{equation}
yielding the time elapsed during the evolution of the information from state $\mathcal{I}_0$ to $\mathcal{I}$. This solution suggests that for $K=\mathcal{C}=0$ the elapsed time is infinite, which is nothing else but another expression for that the information does not change but is conserved. Otherwise, for $K\to\infty$ the system seems to evolve at a diverging rate. However, inspection of the second last equation reveals that d$t=0$ in this case, and the entire expression becomes obsolete. This case corresponds to complete stochasticity with no production nor evolution of information at all, the final  thermodynamic equilibrium state of maximum information.

The important case is the intermediate one where $K$ is finite, corresponding to a state of nonlinear non-stochastic (chaotic)  interactions. They cause the information to evolve in finite time. Speculatively, this expression can be taken as an equation for the ``production of time". In such an interpretation, time is ``generated" under circumstances when information is produced - a physically not unreasonable assertion. In this interpretation there is more room left for speculation. The case $K=0$ corresponds to stationarity with no time evolution of information. Information can only be redistributed convectively then. $K=\infty$ means complete stochasticity. Hence, stochastic systems do, in this interpretation, not produce any time at all. Time, again in this interpretation, is attributed to the nonlinear, non-stochastic action that generates information in a complex system.

Though these remarks are intriguing, the sound physical result is contained in the one-subsystem (particle) kinetic equation of information which forms the basic equation for the evolution of information. We have not given an explicit derivation of its right hand side, the correlation functional $\mathcal{C}=K$, the  Kolmogorov entropy rate. This is left as an exercise for future research until the hierarchy equations have been constructed. Rigorous construction of the hierarchy equations is inhibited in this communication by restriction of space. {We have also neither invested effort into any quantum mechanical nor field theoretical formulation which both are of utmost interest in applications, nor have we envisaged investigation of any different more general definition of information than the classical Boltzmann-Shannon information. }











\end{document}